\documentstyle[12pt]{article}
\begin{document}
\title{Weak radiative hyperon decays, Hara's theorem and the diquark}
\author{
{P. \.{Z}enczykowski}$^*$\\
\\
{\em Dept. of Theor. Physics} \\
{\em Institute of Nuclear Physics}\\
{\em Radzikowskiego 152,
31-342 Krak\'ow, Poland}\\
}
\maketitle
\begin{abstract}
Weak radiative hyperon decays are discussed in the diquark-level approach.
It is pointed out that in the general diquark formalism one 
may reproduce the experimentally suggested pattern
of asymmetries, while maintaining
Hara's theorem in the SU(3) limit.
At present, however,  
no detailed quark-based model of parity-violating diquark-photon 
coupling exists that
would have the necessary properties.
\end{abstract}
$^*$ E-mail:zenczyko@solaris.ifj.edu.pl\\
\vskip 0.8in
\begin{center}REPORT \# 1767/PH, INP-Krak\'ow \end{center}
\newpage

Weak radiative hyperon decays (WRHD's) have
proven to be a challenge to our theoretical
understanding.  Despite many years of theoretical studies, a
satisfactory description of these processes is still lacking.
For a review see ref.\cite{LZ} where current
theoretical and experimental situation in the field is presented.

The puzzle posed by weak radiative hyperon decays contains a couple of
ingredients, most of which relate to the issue of Hara's theorem \cite{Hara}. 
This theorem, originally formulated at the hadron level,
states that the parity-violating amplitude of decay
$\Sigma ^+ \rightarrow p \gamma $ should vanish in the limit of SU(3)
flavour symmetry. Since SU(3) symmetry is expected to be weakly broken,
the parity-violating amplitude in question and, consequently, 
the $\Sigma ^+ \rightarrow p \gamma $ decay asymmetry should be small. 
Experiment \cite{Foucher} shows, however, that the asymmetry is large:
\begin{equation}
\label{eq:Foucher}
\alpha (\Sigma ^+ \rightarrow p \gamma )= -0.72 \pm 0.086 \pm 0.045
\end{equation}
Satisfactory explanation of such a large value of this asymmetry
constitutes a theoretical problem, which is even more difficult
when one demands a successful simultaneous description of the
experimental values of the asymmetries of three related WRHD's, namely
$\Lambda \rightarrow n \gamma $, $\Xi ^0 \rightarrow \Lambda \gamma $, 
and $\Xi ^0 \rightarrow \Sigma ^0 \gamma $.

Theoretical calculations may be divided into those 
ultimately carried out at hadron level (eg. \cite{Gavela})
and those performed totally at quark level (eg. \cite{KR,VS}).
Hadron-level
calculations such as those of Gavela et al. \cite{Gavela} 
do satisfy Hara's theorem in the SU(3) limit 
and may yield fairly large negative $\Sigma ^+ \rightarrow p \gamma $
asymmetry but have problems with a simultaneous
description of all four asymmetries \cite{LZ} which are predicted to be 
negative.  The data, on the other hand, seem to indicate that the
$\Xi ^0 \rightarrow \Lambda \gamma$ asymmetry is positive.
Strict quark-level calculations and the hadron-level VMD approach of 
ref.\cite{Zen}
describe the whole body of data significantly better \cite{LZ}. 
However, they achieve this at the cost of apparently violating Hara's 
theorem in the SU(3) limit.
Unless some other mechanism or effect intervenes, it seems therefore
that the data indicate that Hara's theorem is broken. 

Very small size of the
experimental $\Xi ^- \rightarrow \Sigma ^- \gamma $ 
branching ratio \cite{Dubbs} relative to those of the
$\Sigma ^+ \rightarrow p \gamma $, 
$\Lambda \rightarrow n \gamma $, $\Xi ^0 \rightarrow \Lambda \gamma $, 
and $\Xi ^0 \rightarrow \Sigma ^0 \gamma $ processes
means that 
$s \rightarrow d \gamma $ single-quark transitions are negligible 
in $\Sigma ^+$, $\Lambda $, and $\Xi ^0 $ decays. 
Thus, as already indicated by the analysis of Gilman and Wise \cite{GW},
the dominant contribution must come from processes
involving at least two quarks.
In both hadron- and quark-level approaches it is
the $W$-exchange between two quarks 
(one of which emits a photon)
that provides the dynamics underlying such two-quark processes.
Both types of existing approaches use a more or less complicated prescription
for the calculation of the relevant amplitudes. 
However, the basic transition is essentially a {\em diquark} $\rightarrow $ 
{\em diquark} + $\gamma $ process,
and thus a simple description of amplitudes
(essentially a counterpart of model-independent single-quark
analysis of Gilman and Wise) should be possible.
It is the purpose of this note to fill in this gap and to analyse the
diquark transitions in a manner somewhat similar to GW analysis, 
while attempting to maintain Hara's theorem in the SU(3) limit. 

To begin with, let us observe that in the $\Sigma ^+$, $\Lambda $, and 
$\Xi ^0$ radiative decays under consideration
the basic {\em diquark} $\rightarrow $ 
{\em diquark} + $\gamma $ transition is 
everywhere \underline{the same} $(us) \rightarrow (ud) + \gamma $
process. The $(us)$ or $(ud)$ diquark may be in spin 0 or spin 1 state. 
Let us therefore denote possible diquark states by 
$A(q_1q_2)$ and $S(q_1,q_2)$ ($q_1q_2=ud$ or $us$):
\begin{eqnarray}
\label{eq:diquarks}
|A(q_1q_2)\rangle &=&
|\frac{1}{\sqrt{2}}(q_1q_2-q_2q_1)
\frac{1}{\sqrt{2}}(\uparrow \downarrow-\downarrow \uparrow)\rangle \nonumber \\
|S^{+1}(q_1q_2)\rangle &=& 
|\frac{1}{\sqrt{2}}(q_1q_2+q_2q_1) \uparrow \uparrow \rangle
\nonumber \\
|S^{0}(q_1q_2)\rangle &=& 
|\frac{1}{\sqrt{2}}(q_1q_2+q_2q_1)
\frac{1}{\sqrt{2}}(\uparrow \downarrow+\downarrow \uparrow)\rangle \nonumber \\
|S^{-1}(q_1q_2)\rangle &=&
|\frac{1}{\sqrt{2}}(q_1q_2+q_2q_1) \downarrow \downarrow \rangle
\end{eqnarray}

We now  rewrite the SU(6) wave functions of relevant initial 
and final
baryons (containing $us$ and $ud$ diquarks respectively) in terms of
a diquark and a spectator quark:
\begin{eqnarray}
\label{eq:wavefunctions}
\langle p\uparrow| &=&
\frac{1}{\sqrt{2}}\langle A(ud)|\langle u\uparrow | + 
\frac{1}{3\sqrt{2}}\langle S^{0}(ud)|\langle u\uparrow |
-\frac{1}{3}\langle S^{+1}(ud)|\langle u\downarrow | + ...\nonumber \\
|\Sigma ^+ \downarrow \rangle &=&
-\frac{1}{\sqrt{2}}|A(us)\rangle |u\downarrow \rangle + 
\frac{1}{3\sqrt{2}}|S^{0}(us)\rangle |u\downarrow \rangle
-\frac{1}{3}|S^{-1}(us)\rangle |u\uparrow \rangle + ...\nonumber \\
\langle n\uparrow| &=&
-\frac{1}{\sqrt{2}}\langle A(ud)|\langle d\uparrow | + 
\frac{1}{3\sqrt{2}}\langle S^{0}(ud)|\langle d\uparrow |
-\frac{1}{3}\langle S^{+1}(ud)|\langle d\downarrow | + ...\nonumber \\
|\Lambda \downarrow \rangle &=&
\frac{1}{2\sqrt{3}}|A(us)\rangle |d\downarrow \rangle +
\frac{1}{2\sqrt{3}}|S^{0}(us)\rangle |d\downarrow \rangle -
\frac{1}{\sqrt{6}}|S^{-1}(us)\rangle |d\uparrow \rangle + ...\nonumber \\
\langle \Lambda \uparrow| &=&
-\frac{1}{\sqrt{3}}\langle A(ud)|\langle s\uparrow | + ...\nonumber \\
|\Xi ^0 \downarrow \rangle&=&
\frac{1}{\sqrt{2}}|A(us)\rangle|s\downarrow \rangle +
\frac{1}{3\sqrt{2}}|S^{0}(us)\rangle |s\downarrow \rangle -
\frac{1}{3}|S^{-1}(us)\rangle |s\uparrow \rangle + ... \nonumber \\
\langle \Sigma ^{0} \uparrow | &=&
\frac{\sqrt{2}}{3}\langle S^{+1}(ud)|\langle s\downarrow | -
\frac{1}{3} \langle S^{0}(ud)|\langle s\uparrow | + ...
\end{eqnarray}

In Eq.(\ref{eq:wavefunctions}) we have explicitly written down only those
diquarks within which the weak $us \rightarrow ud $ transition may take place.
Denoting weak + electromagnetic {\em diquark} $\rightarrow$ {\em diquark}
+ $\gamma $ transition amplitudes by 
\begin{eqnarray}
\label{eq:diquarktransitions}
t_{+1}&=&\langle S^{+1}(ud)\gamma|T|A(us)\rangle\nonumber \\
t_{-1}&=&\langle A(ud) \gamma|T|S^{-1}(us)\rangle\nonumber \\
v&=&\langle S^{0}(ud)\gamma|T|S^{-1}(us)\rangle + 
\langle S^{+1}(ud)\gamma |T|S^{0}(us)\rangle
\end{eqnarray}
(the momenta of final baryon and photon 
define the axis of spin quantisation)
and using Eqs.(\ref{eq:wavefunctions}), 
we may express the amplitudes of WRHD's in terms of $t_{+1}$, $t_{-1}$, and $v$:
\begin{eqnarray}
\label{eq:amplitudes}
\langle p\uparrow \gamma |T|\Sigma ^+\downarrow \rangle &=&
\frac{1}{3\sqrt{2}}\;t_{+1} - \frac{1}{3\sqrt{2}}\;t_{-1}-
\frac{1}{9\sqrt{2}}\;v \nonumber \\
\langle n\uparrow \gamma |T|\Lambda \downarrow\rangle &=&
-\frac{1}{6\sqrt{3}}\;t_{+1} + \frac{1}{2\sqrt{3}}\;t_{-1}-
\frac{1}{6\sqrt{3}}\;v \nonumber \\
\langle \Lambda \uparrow \gamma |T| \Xi ^0 \downarrow \rangle &=&
\frac{1}{3\sqrt{3}}\;t_{-1} \nonumber \\
\langle \Sigma ^0 \uparrow \gamma |T| \Xi ^0 \downarrow \rangle &=&
\frac{1}{3}\;t_{+1} + \frac{1}{9}\;v
\end{eqnarray}
The above formulas are valid both for parity-violating and for parity-conserving
amplitudes (with different parameters), for \underline{any} two-quark 
$us \rightarrow ud + \gamma$ processes. 
With six parameters it should not be difficult
to fit the four experimental asymmetries and four branching ratios 
if the two-quark transitions are indeed dominant.
We have not attempted such a fit since
1) the experimental numbers still carry quite significant errors,
and 2) we are more interested in the
theoretical problem which manifests itself in parity-violating amplitudes
only.
From now on we will accept that the parity-conserving amplitudes are well
described by the standard pole model prescription (eg. \cite{Gavela}).
Thus, only three parameters are at our disposal.  Please note also
that the $\Xi ^0 \rightarrow \Lambda \gamma $ asymmetry is especially
interesting as it provides a direct measure of a single diquark
amplitude $t_{-1}$.

For the sake of a subsequent 
discussion let us restrict the meaning of Eq.(\ref{eq:amplitudes}) to
the parity-violating sector and let us re-express the amplitudes
in terms of three parameters $P$, $Q$ and $v$, where $P$ and $Q$ are given by:
\begin{eqnarray}
\label{eq:parameters}
P&=&\frac{1}{2}(t_{+1}+t_{-1})-\frac{1}{6}v\nonumber \\
Q&=&\frac{1}{2}(t_{+1}-t_{-1})-\frac{1}{6}v
\end{eqnarray}

The amplitudes in question expressed in this way are given in the second
column of Table 1.
Phase convention used in our formulas 
is such that the signs of parity-violating amplitudes in Table 1 are
automatically equal to the sign of asymmetries
once the common sign of parity-conserving amplitudes is fixed as positive.
From Table 1 it can be seen that pole and quark models correspond to
a different choice of diquark parameters $P$, $Q$, and $v$.
For the pole model \cite{Gavela} 
we have $P=C$ and $Q=Cx$ with $C=1/(1-x^2)$ (see ref.\cite{LZ}),
where $x$ is the SU(3) breaking parameter estimated in \cite{LZ} to be 
$x=\delta s/\Delta \omega \approx 1/3$
($\delta s = m_s-m_d \approx 190 MeV$, $\Delta \omega = m(1/2^-)-m(1/2^+)
\approx 570 MeV$). In the pole model, 
Hara's theorem is satisfied 
in the SU(3) limit $x \rightarrow 0$
and, consequently, in this limit only the $P$ parameter is nonzero.
For the quark model/VDM approach \cite{VS,Zen}
$P=Cx$ and $Q=C$. If formulas of these approaches really describe
parity-violating amplitudes in full, one obtains violation of
Hara's theorem as $Q$ does not vanish in the SU(3) limit.
Since explicit calculations of diquark-photon couplings
in pole or quark models yield either $v=0$ or $v\approx 0$, we will
neglect $v$ in the following. For an explicit calculation of $v$ in the 
constituent quark
model of a diquark see ref.\cite{UV}, where a very small $v$ proportional to
$(\delta s/m_d)^2$ is obtained.

To proceed let us consider now the most general 
gauge-invariant {\em diquark} $\rightarrow $
{\em diquark} + $\gamma $ parity-violating strangeness-changing interaction:
\begin{equation}
\label{eq:interaction}
{\cal L} \propto (J_{\mu } + J_{\mu}^{\dagger})\;A^{\mu }
\end{equation}

\noindent 
with $J_{\mu}$ being the strangeness-changing 
$0^+ \rightarrow 1^+$ diquark current 
\begin{equation}
\label{eq:diquarkcurrent}
J^{\mu} = g(q^2)\; [(q\cdot k) \epsilon_S^{\mu *} - p^{\mu}\epsilon_S^*\cdot q]
\end{equation}
where $\epsilon ^*_S$ describes polarization of the final $1^+$ diquark,
$g$ is a real function of $q^2$,
and $p$, $k$, and $q$ are momenta of the initial diquark, final diquark
and photon respectively.
$J^{\mu \dagger}$ is obtained from Eq.(\ref{eq:diquarkcurrent}) by changing
$q \rightarrow -q$, $\epsilon ^*_S \rightarrow \epsilon _S $, while leaving $p$
and $k$ unchanged (together with their interpretation of initial and final
diquark momenta).
For real transverse photons there will be no contribution from the $p^{\mu}$
term since, upon integration over diquark momentum in the initial baryon,
the terms of opposite $p^k$ will cancel.  

Using the diquark-photon interaction of Eqs.(\ref{eq:interaction},
\ref{eq:diquarkcurrent}) we find that the {\em diquark}
$\rightarrow $ {\em diquark} + $\gamma $ transition amplitudes are proportional
to:
\begin{eqnarray}
\label{eq:t1t-1}
t_{+1}&\propto&g(0)\;[m^2(A(u,s))-m^2(S(u,d))]\nonumber \\
t_{-1}&\propto&g(0)\;[m^2(A(u,d))-m^2(S(u,s))]
\end{eqnarray}
where $m(A(u,q))$, $m(S(u,q))$ are masses of spin 0 and spin 1 diquarks.

Let us now assume that diquarks composed of $us$ are heavier than those
made of $ud$:
\begin{eqnarray}
\label{eq:SU3breaking}
m^2(A(u,s))&=&m^2(A(u,d))+\Delta s\nonumber \\
m^2(S(u,s))&=&m^2(S(u,d))+\Delta s
\end{eqnarray}

\noindent
with $\Delta s$ vanishing in the SU(3) limit.  We then find
\begin{eqnarray}
\label{eq:PQ}
P&\propto&g(0)\;[m^2(A(u,d))-m^2(S(u,d))]\nonumber \\
Q&\propto&g(0)\;\Delta s
\end{eqnarray}

Ways of breaking SU(3) symmetry that are different in detail from the simple
version of Eq.(\ref{eq:SU3breaking}) may be also considered.
As long as in the SU(3) limit one has $m(A(u,s))\rightarrow m(A(u,d))$ and 
$m(S(u,s))\rightarrow m(S(u,d))$, the qualitative results
of our discussion will not change.
One has to remember, however, that if function $g$  also
depends on diquark masses, this might cancel the $m$-dependence of 
Eq.(\ref{eq:t1t-1}).  For example, in the VMD-based approach
to parity-violating amplitudes in $B\rightarrow K^* \gamma $ decay 
(analogous to our diquark decay), Golowich
and Pakvasa \cite{GP} use the gauge-invariant coupling $\epsilon _{\mu }(\gamma )
J^{\mu } = g'\epsilon _{\mu }(\gamma )(\epsilon ^{\mu *}(K^*)
- \frac{1}{q\cdot p} \epsilon ^*(K^*) \cdot q \;p^{\mu }) $.

Upon inspection of Table 1 we see that in the gauge-invariant diquark-level 
approach Hara's theorem is recovered in the
SU(3) limit ($Q \rightarrow 0$). 
In addition we see that one may obtain $P \approx 0$
if masses of $A$ and $S$ diquarks are similar, ie. in the spin symmetry  limit.
Thus, it is possible to obtain the 
signature (--,+,+,--) of  the
$\Sigma ^+ \rightarrow p \gamma $, 
$\Lambda \rightarrow n \gamma $, $\Xi ^0 \rightarrow \Lambda \gamma $, 
and $\Xi ^0 \rightarrow \Sigma ^0 \gamma $
asymmetries 
(characteristic of the quark model/VDM approach
and also suggested by experiment),
and yet maintain Hara's theorem in the SU(3) limit.
In other words, large asymmetries with signature (--,+,+,--) are 
compatible with Hara's theorem, provided the
SU(3)-breaking term is much larger than the SU(3)-symmetric term
($Q \gg P$).

The diquark approach of this paper seems to suggest 
that although $Q\rightarrow 0$ in the SU(3) limit,
it may happen that $Q \gg P$ for realistic SU(3) breaking.
However, the problem remains how to achieve this
in a microscopic model of the diquark.
Explicit calculations in the pole model \cite{Gavela,LZ} give just
the opposite: $Q/P \approx 1/3$. One can obtain $Q \gg P$ in the pole
model provided that $x = \delta s/\Delta \omega \gg 1$.  This corresponds
to the Li-Liu proposal \cite{LL} 
which was opposed by Gaillard \cite{MKG}. 

On the other hand, explicit calculations at quark level 
(such an approach to diquark lies at the
basis of ref.\cite{UV})
yield in the SU(3) limit 
\begin{equation}
\label{eq:explicitdq}
t_{+1}=-t_{-1}\ne 0
\end{equation}
leading therefore to an apparent
violation of Hara's theorem. Technically, the origin of the above
result is obviously the same as in the quark model calculations of 
Kamal and Riazuddin \cite{KR}. 
The diquark approach of ref.\cite{UV}
is conceptually identical to the quark-level approach of \cite{KR,VS}.

Recently it was argued \cite{Azimov} that 
the $\overline{\psi}_{p}\gamma_5\gamma_{\mu}\psi _{\Sigma ^+}A^{\mu}$ 
term, to the appearance of 
which the quark-model Hara's-theorem-violating results for
$\Sigma ^+ \rightarrow p \gamma $ were assigned, may be renormalized away. 
It would be interesting to study if and how such a 
renormalization procedure
affects quark model predictions for the asymmetries of
four related WRHD's ($\Sigma ^+ 
\rightarrow p \gamma$, $\Lambda \rightarrow n \gamma$,
$\Xi ^0 \rightarrow \Lambda \gamma$, 
and $\Xi ^0 \rightarrow \Sigma ^0 \gamma$),
and how it compares with the general diquark-level approach
which provides an after-renormalization description.

In summary, the diquark-level gauge-invariant approach to parity-violating 
amplitudes of WRHD's is in principle capable of explaining the experimentally
suggested
pattern of asymmetries without violating Hara's theorem in the SU(3) limit.
However, in order to achieve this one needs large SU(3)-breaking
and small SU(3)-symmetric terms in parity-violating diquark-photon couplings.
At present no existing quark-based model of a diquark exhibits this property.

ACKNOWLEDGEMENTS.

This work was partially supported by the
KBN grant No 2P03B23108.

\newpage
Table 1

Parity-violating amplitudes in the general diquark approach, and
the pole and quark models.  Expressions in the last
two columns are taken from ref.\cite{LZ} with $x$ being the SU(3)
breaking parameter and $C = 1/(1-x^2)$.

\noindent
\begin{tabular}{|l|c|c|c|}
\hline
process &diquark&pole model&VDM/quark model\\
\hline
$\Sigma ^+ \rightarrow p \gamma $     &$-\frac{2}{3\sqrt{2}}Q$  &
$-\frac{2}{3\sqrt{2}}Cx$  &$-\frac{2}{3\sqrt{2}}C$\\
$\Lambda \rightarrow n \gamma $       
&$-\frac{1}{3\sqrt{3}}P + \frac{2}{3\sqrt{3}}Q+\frac{2}{9\sqrt{3}}v$  
&$-\frac{1}{3\sqrt{3}}C + \frac{2}{3\sqrt{3}}Cx$
&$-\frac{1}{3\sqrt{3}}Cx + \frac{2}{3\sqrt{3}}C$\\
$\Xi ^0 \rightarrow \Lambda \gamma $  
&$-\frac{1}{3\sqrt{3}}P + \frac{1}{3\sqrt{3}}Q$  
&$-\frac{1}{3\sqrt{3}}C + \frac{1}{3\sqrt{3}}Cx$
&$-\frac{1}{3\sqrt{3}}Cx + \frac{1}{3\sqrt{3}}C$\\
$\Xi ^0 \rightarrow \Sigma ^0 \gamma $
&$-\frac{1}{3}P - \frac{1}{3}Q -\frac{2}{9}v$  
&$-\frac{1}{3}C - \frac{1}{3}Cx$
&$-\frac{1}{3}Cx - \frac{1}{3}C$\\
\hline
\end{tabular}

\end{document}